\overfullrule=0pt
\input harvmac
\def\a{{\alpha}}
\def\L{{\Lambda}}

\def\b{{\beta}}

\def\g{{\gamma}}

\def\d{{\delta}}

\def\e{{\epsilon}}

\def\half{{1\over 2}}
\def\p{{\partial}}

\Title{\vbox{\hbox{~ }}}
{\vbox{
\centerline{\bf Constrained BV Description of String Field Theory}}}
\bigskip\centerline{Nathan Berkovits\foot{e-mail: nberkovi@ift.unesp.br}
}
\bigskip
\centerline{\it ICTP South American Institute for Fundamental Research}
\centerline{\it Instituto de F\'\i sica Te\'orica, UNESP-Universidade Estadual
Paulista}
\centerline{\it R. Dr. Bento T. Ferraz 271 - Bl. II, 01140-070, S\~ao Paulo, SP, Brasil}

\vskip .3in
In the conventional BV description of string field theory, the string field $\Phi$
is split as $\Phi = \Psi +\Psi^*$ where $\Psi$ includes all states with ghost number $\leq G$
and describes the spacetime fields, and $\Psi^*$ includes all states with ghost number $> G$
and describes the spacetime antifields. A new approach is proposed here in which separate string
fields $\Psi$ and $\Psi^*$ of unrestricted ghost number describe the spacetime fields and antifields.
The string antifield $\Psi^*$ is constrained to satisfy $\Psi^* = {\partial L}/{\partial(Q\Psi)}$
where $L$ is the BV Lagrangian and $Q$ is the worldsheet BRST operator. Dirac antibrackets are defined
using this constraint, and the resulting description is equivalent to the conventional BV description
for open and closed bosonic string field theory. For open superstring field theory, this constrained
BV description is much simpler than the conventional BV description and allows the BV action to be
expressed in the same WZW-like form as the classical action.

\Date{January 2012}

\newsec{Introduction}

Field theory actions with reducible gauge symmetries require
ghosts and ghosts-for-ghosts. A convenient formalism for describing 
such actions is the Batalin-Vilkowisky (BV) formalism \ref\batalin{
I. A. Batalin and G. A. Vilkovisky, ``Quantization of gauge theories with linearly dependent generators'', Phys. Rev. D28 (1983) 2567.} in which the
action is expressed in terms of fields $\psi^I$ and antifields $\psi_I^*$.
The fields $\psi^I$ typically describe the physical states as well as the 
Faddeev-Popov ghosts
and ghosts-for-ghosts. And for each field $\psi^I$, one introduces an antifield
$\psi^*_I$ with opposite statistics satisfying the BV antibracket 
\eqn\brackone{\{\psi^*_I, \psi^J \} = \d_I^J.}
Note that the antibracket is fermionic since $\psi^*_I$ and $\psi^J$ have opposite
statistics but $\d^J_I$ is bosonic.

This BV description closely resembles classical mechanics in which the fields $\psi^I$
are identified with coordinates and the antifields $\psi^*_I$ are identified with momenta. However, since fields and
antifields have opposite statistics, the ``time" derivative $\p\over\p t$ which relates coordinate and momenta should be a fermionic
operation. It has been previously proposed \ref\bering{I.A. Batalin, K. Bering, P.H. Damgaard, ``Superfield Quantization",
Nucl.Phys. B515 (1998) 455-487, arXiv:hep-th/9708140.}
that the BRST operator plays the role of such a fermionic time derivative, and this
proposal will be confirmed here for string field theory \foot{I would like to thank Klaus Bering for describing to me this previous
proposal.}. Just as momenta $p_I$ in classical mechanics are defined by the constraint
$p_I = {\p L(q, {{\p q}\over{\p t}})}/{\p({{\p q^I}\over{\p t}})}$, it will be shown that antifields in string field theory can be defined
by the constraint
\eqn\cons{\Psi^* = {\p L(\Psi, Q\Psi)}/{\p(Q\Psi)}}
where $Q$ is the worldsheet BRST operator and $L$ is the string field theory Lagrangian.
This new interpretation of antifields will reproduce the conventional BV description of open and closed
bosonic string field theory and will drastically simplify the BV description of open superstring
field theory.

The spacetime field theory action for string theory has a complicated set of reducible
gauge symmetries, and the BV formalism is an extremely efficient method for organizing the ghost structure \ref\sieg{
W. Siegel, {\it Introduction to String Field Theory}, World Scientiﬁc, Singapore, (1988), hep-th/0107094. }\ref\witb{
E. Witten, ``Noncommutative geometry and string field theory'', Nucl. Phys. B268 (1986) 
253.} 
\ref\thorn{
C. B. Thorn, ``Perturbation theory for quantized string fields'', 
Nucl. Phys. B 287, 61 (1987). }
\ref\boch{M. Bochicchio, ``Gauge fixing for the field theory
of the bosonic string'', Phys. Lett. B193 (1987) 31.}
\ref\zw{B. Zwiebach, ``Closed string field theory: Quantum action and the B-V master equation'',
Nucl.Phys. B390 (1993) 33, hep-th/9206084.}.
In open bosonic string field theory, the classical action is \witb
\eqn\openaction{S = \langle \Phi Q \Phi + {2\over 3}\Phi^3 \rangle}
where $\Phi$ is a fermionic string field of $+1$ worldsheet ghost number, $Q$ is the worldsheet BRST operator
of $+1$ ghost number, and $\langle~\rangle$ denotes the measure factor on a disk of $+3$ ghost number.
In this case, the complete BV action including all ghosts-for-ghosts is obtained by simply
allowing the string field $\Phi$ to have arbitrary ghost number.

In the conventional BV description \boch, 
states of ghost number $\leq 1$ are identified as spacetime fields and states of
ghost number $>1$ are identified as spacetime antifields. So $\Phi$ splits as 
\eqn\split{\Phi = \Psi + \Psi^*}
where $\Psi$ only includes states with ghost number $\leq 1$ and describes spacetime fields, and
$\Psi^*$ only includes states with ghost number $> 1$ and describes spacetime antifields. 
The BV antibracket of \brackone\ then implies that 
\eqn\brackPhione{\{\Phi(Y), \Phi(Y')\} = \d (Y-Y')}
where $Y$ includes all modes of all worldsheet variables and $\d(Y-Y')$ is defined such that
$\langle \Phi(Y) \d(Y-Y')\rangle = \Phi(Y')$. 

In the new constrained BV description, $\Phi$ of \openaction\ will be replaced by a string field $\Psi$ with unrestricted
ghost number which describes the spacetime fields.  
The spacetime antifields will then be described by a separate string field $\Psi^*$ with unrestricted
ghost number which is required to satisfy the constraint
of \cons\
where $L$ is the Lagrangian of \openaction.
Introducing separate string fields $\Psi$ and $\Psi^*$ with unrestricted ghost number naively 
doubles the number of fields and antifields. But the constraint of \cons\ implies that
$\Psi^* = {\p L}/{\p(Q\Psi)} = \Psi$
which cuts by half the
number of independent fields and antifields. Note that the constraint $\Psi^*-\Psi=0$ is second-class, which means the BV
antibrackets of \brackone\
need to be modified using the standard Dirac procedure as in \ref\beringd{
I. A. Batalin, I. V. Tyutin,
``On possible generalizations of field-antifield formalism'',
Int.J.Mod.Phys.A8:2333-2350 (1993), hep-th/9211096.}.
Following this Dirac procedure, one reproduces
the antibracket of \brackPhione\ which proves that the constrained and conventional BV descriptions of open bosonic string
field theory are equivalent.

This Dirac procedure for the kinetic
term $\langle \Phi Q\Phi\rangle$ is similar to the quantization procedure for 
a Weyl spinor $\psi^\a$ in field theory. The spinor kinetic
term $S=\int d^D x ~\psi^\a \g^n_{\a\b} \p_n \psi^\b$ 
implies that the canonical momenta for
$\psi^\alpha$ is $\psi^*_\alpha = \p L/\p(\p_0\psi^\a) = (\g^0 \psi)_\alpha$. The constraint $\psi^*_\a - 
(\g^0\psi)_\a =0$ is a second-class constraint, and the resulting Dirac bracket is
$\{\psi^\a (x), \psi^\b (x')\} =\half \g_0^{\a\b} \d^D(x-x')$.

Similarly, in closed bosonic string field theory, the classical action is \zw
\eqn\closedact{
S= \langle \Phi Q (c-\bar c)_0 \Phi \rangle + ...}
where $\Phi$ is a bosonic string field of $+2$ ghost number satisfying the constraint $(b-\bar b)_0 \Phi=0$, $Q$ is the
sum of the left and right-moving BRST operators, $(b_0 ,c_0)$ and $(\bar b_0,\bar c_0)$ are the zero modes
of the left and right-moving Virasoro ghosts, $\langle ~\rangle$ denotes the measure factor on a sphere
of $+6$ ghost number, and $...$ denote the interaction terms.
As before, the complete BV action including all ghosts-for-ghosts is obtained by simply
allowing the string field $\Phi$ to have arbitrary ghost number. 

In the conventional BV description \zw, states of ghost number $\leq 2$ in $\Phi$ are identified as spacetime fields and states of
ghost number $>2$ are identified as spacetime antifields. 
So $\Phi$ splits as 
\eqn\split{\Phi = \Psi + \Psi^*}
where $\Psi$ only includes states with ghost number $\leq 2$ and describes spacetime fields, and
$\Psi^*$ only includes states with ghost number $> 2$ and describes spacetime antifields. 
The BV antibracket of \brackone\ then implies that 
\eqn\brackPhione{\{\Phi(Y), \Phi(Y')\} = (b-\bar b)_0 \d (Y-Y')}
where the factor of $(b_0-\bar b_0)$ comes from the constraint on $\Phi$.

In the new constrained description, $\Phi$ of \closedact\ is  
replaced by $\Phi = (b-\bar b)_0 \Psi$ where $\Psi$ is a fermionic string field of unrestricted ghost number which
describes spacetime fields. In terms of $\Psi$, the action of \closedact\ is 
\eqn\closedtwoact{
S= \langle \Psi Q (b-\bar b)_0 \Psi \rangle + ...,}
so the bosonic string field $\Psi^*$ for the spacetime antifields is defined by
\eqn\closedanti{\Psi^* = \p L/\p(Q\Psi) = (b-\bar b)_0 \Psi.}
In this case, the constraint $\Psi^* - (b-\bar b)_0\Psi=0$ has a first-class and second class
piece where the first-class piece is $(b-\bar b)_0 \Psi^*=0$ and generates the gauge invariance 
\eqn\gaugeclosed{\d\Psi = (b-\bar b)_0 \Lambda}
for any $\Lambda$. One can then use the standard Dirac procedure to compute the antibracket of any
operators which are gauge-invariant under \gaugeclosed. The operator $\Phi=(b_0-\bar b_0)\Psi$ is gauge invariant,
and one can easily show that its antibracket reproduces the conventional result of \brackPhione.

Finally, in the WZW-like version of open superstring field theory, the conventional BV construction is
much more complicated. The classical action for the Neveu-Schwarz sector is 
\ref\nb{N. Berkovits, ``Super-Poincare invariant superstring field theory'',
Nucl.Phys. B450 (1995) 90, hep-th/9503099.}
\eqn\nsaction{S = \langle (e^{-\Phi} Q e^\Phi)(e^{-\Phi}\eta_0 e^\Phi) +\int_0^1 dt
(e^{-t\Phi} \p_t e^{t\Phi})\{(e^{-t\Phi} Q e^{t\Phi}),(e^{-t\Phi}\eta_0 e^{t\Phi})\} \rangle}
where $\Phi$ is a string field in the ``large'' Hilbert space with zero picture and zero
ghost number, $Q$ is the worldsheet BRST operator, $\eta_0$ carries $-1$ picture and $+1$ ghost
number and is the zero mode coming from Friedan-Martinec-Shenker
fermionization of the $(\beta,\gamma)$ ghosts,
and $\langle~\rangle$
is the standard measure factor in the ``large'' Hilbert space which carries $-1$ picture and $+2$ ghost number.
This classical action has the WZW-like gauge invariances 
\eqn\wzwgauge{\d (e^\Phi) = (Q \Lambda_{-1,0}) e^\Phi + e^\Phi (\eta_0 \Lambda_{-1,1})}
where $\Lambda_{g,p}$ are gauge parameters with ghost number $g$ and picture $p$.

As explained in \ref\free{
M.~Kroyter, Y.~Okawa, M.~Schnabl, S.~Torii and B.~Zwiebach,
``Open superstring field theory I: gauge-fixing, ghost structure and propagator'', to appear today.}
\ref\us{N.~Berkovits, M.~Kroyter, Y.~Okawa, M.~Schnabl, S.~Torii and B.~Zwiebach,
``Open superstring field theory II: approaches to the BV master action'',
in preparation.}
\ref\torii{S. Torii, ``Validity of Gauge-Fixing Conditions and the Structure of Propagators
 in Open Superstring Field Theory,'' to appear today.}
\ref\torii2{
S. Torii, ``Gauge Fixing of Open Superstring Field Theory in the Berkovits Non-polynomial Formulation,''
Prog. Theor. Phys. Suppl. 188 (2011) 272.}, the conventional BV description of this action
involves an infinite pyramid of ghosts-for-ghosts $\Phi_{g,p}$ together with their antifields. 
This can be seen from the 
linearized version of the above gauge transformation 
\eqn\lingauge{\d\Phi = Q\Lambda_{-1,0} + \eta_0\Lambda_{-1,1},}
which has the linearized gauge-for-gauge invariance
\eqn\lingaugetwo{\d\Lambda_{-1,0} =Q\Lambda_{-2,0} + \eta_0\Lambda_{-2,1},
\quad \d\Lambda_{-1,1} = Q \Lambda_{-2,1} + \eta_0\Lambda_{-2,2},}
which has linearized gauge-for-gauge-for-gauge invariances, etc. Generalizing the classical action of \nsaction\ to
include this infinite pyramid of ghosts-for-ghosts and their antifields is a difficult problem since it involves
states of different pictures which interact in a complicated manner.
Partial results for the construction of this nonlinear conventional BV action will be described in \us.

On the other hand, the open superstring field theory action easily generalizes to include
ghosts-for-ghosts using the constrained BV description. One simply replaces the string field $\Phi$ in \nsaction\ with
a string field $\Psi$ of zero picture and unrestricted ghost number which describes the spacetime fields.
The spacetime antifields are described by a fermionic string field $\Psi^*$ of $-1$ picture and unrestricted ghost number
which is constrained to satisfy 
\eqn\superanti{\Psi^* = \p L/\p(Q\Psi) = \eta_0\Psi + {1\over 3}(\Psi(\eta_0\Psi) - (\eta_0\Psi)\Psi) + ... }
where $L$ is the Lagrangian of \nsaction\ and the nonlinear terms in $...$ are easily determined.
As in the case of closed bosonic string field theory, the constraint of \superanti\ has a first-class
and second-class piece where the first-class piece generates the gauge invariance 
\eqn\gaugesuper{\d(e^\Psi) = e^\Psi (\eta_0 \Lambda).}

One can then use the standard Dirac procedure to compute the antibracket of any
operators which are gauge-invariant under \gaugesuper. The operator $J= (\eta_0 e^\Psi) e^{-\Psi}$ is gauge invariant,
and satisfies the antibracket\foot{I would like to thank Andrei Mikhailov for suggesting a simple form for this antibracket.}
\eqn\antiJ{ \{J^K(Y), J^L(Y')\} = h^{KL} \eta_0 \d(Y-Y') + f^{KL}_M J^M(Y) \d(Y-Y')}
where $(K,L,M)$ are Lie-algebra indices coming from the Chan-Paton factors and $h^{KL}$ and
$f^{KL}_M$ are the metric and structure constants of the Lie algebra.
The BV action $S$ of \nsaction\ is also gauge-invariant under \gaugesuper, and one finds that
\eqn\antiS{\{ S, J^K(Y)\} = Q (J^K(Y)), \quad\{S, S\}=0.}
The antibracket of \antiJ\ closely resembles the OPE's of holomorphic Kac-Moody currents of
a two-dimensional WZW sigma model, and the antibracket of \antiS\ implies that the BV transformation
of gauge-invariant operators is equivalent to acting with the worldsheet BRST operator $Q$ which is manifestly nilpotent.

In section 2 of this paper, open bosonic string field theory will be described
using the constrained BV description. In section 3,
closed bosonic string field theory will be described
using the constrained BV description. 
And in section 4, open superstring field theory
will be described using the constrained BV description.

\newsec{Open Bosonic String Field Theory}

\subsec{Conventional BV description}

As discussed in the introduction, the conventional BV action for open bosonic
string field theory is 
obtained from the classical action \witb
\eqn\popenaction{S = \langle \Phi Q \Phi + {2\over 3}\Phi^3 \rangle}
by simply allowing $\Phi$ to have unrestricted ghost number.
In this description, states in $\Phi$ of ghost number $\leq 1$ are identified as spacetime fields $\psi^I$ and states of
ghost number $>1$ are identified as spacetime antifields $\psi^*_I$. So $\Phi$ splits as 
\eqn\ssplit{\Phi = \Psi + \Psi^*}
where $\Psi$ only includes states with ghost number $\leq 1$ and describes spacetime fields $\psi^I$, and
$\Psi^*$ only includes states with ghost number $> 1$ and describes spacetime antifields $\psi^*_I$. 

For general operators $A$ and $B$, the BV antibracket is defined as
\eqn\general{\{A, B\} = 
  (A {\p\over{\p\psi^*_I}}) ({\p\over{\p\psi^I}}B)
- (B {\p\over{\p\psi^*_I}}) ({\p\over{\p\psi^I}} A) }
where $ (A {\p\over{\p\psi^*_I}})$ denotes that the partial derivative ${\p\over{\p\psi^*_I}}$ acts from the right on $A$.
By expanding $\Phi$ in terms of spacetime fields and antifields, this implies that
\eqn\bPhione{\{\Phi(Y), \Phi(Y')\} = \d (Y-Y')}
where $Y$ includes all modes of all worldsheet variables, $\d(Y-Y')$ is defined such that
$\langle \Phi(Y)\d(Y-Y')\rangle_Y = \Phi(Y')$, and $\langle ~\rangle_Y$ denotes functional integration over the $Y$
variables. 
Note that $\d(Y-Y')$ carries ghost number $+3$ and is proportional to 
\eqn\propopen{(c- c')_{-1} (c-c')_0 (c-c')_1}
where $(c_1, c_0, c_{-1})$ are the zero modes on a disk of
the Virasoro $c$ ghost. Since the factor of \propopen\
implies that $\d(Y-Y') = - \d(Y'-Y)$, the definition of \bPhione\ is
consistent with the fact that 
$\{\Phi(Y),\Phi(Y')\} = - \{\Phi(Y'), \Phi(Y)\}.$

Using the action $S$ of \popenaction, one finds that
$S$ satisfies the antibracket
\eqn\sbracket{\{S,\Phi\} =2( Q\Phi + \Phi\Phi )} 
which is interpreted as the BV transformation $\d_{BV}\Phi$. Using the fermionic nature
of the antibracket, one finds that 
\eqn\nilpotent{\d_{BV}\d_{BV}\Phi = \{S, \{S,\Phi\}\} = -2 Q\{S,\Phi\} + 2\{S,\Phi\}\Phi - 2\Phi\{S,\Phi\} }
$$= -4Q^2\Phi - 4Q(\Phi\Phi) +4 (Q\Phi)\Phi +4(\Phi\Phi)\Phi-4 \Phi (Q\Phi)  - 4\Phi (\Phi\Phi) =0.$$
So the BV transformation is nilpotent as desired.

\subsec{Constrained BV description}

In the constrained BV description of open bosonic
string field theory, the BV action is 
\eqn\popenaction{S = \langle \Psi Q \Psi + {2\over 3}\Psi^3 \rangle}
where $\Psi$ has unrestricted ghost number but contains only spacetime fields.
The canonical momenta to $\Psi$ which contains the spacetime antifields is defined by
a separate string field $\Psi^*$ of unrestricted ghost number which is constrained to
satisfy
\eqn\ccons{
\Psi^* = {\p L}/{\p(Q\Psi)} = \Psi.}
Note that both $\Psi$ and $\Psi^*$ are fermionic string fields.

As in classical mechanics, the Poisson antibracket of $\Psi$ and its canonical momentum $\Psi^*$
is defined by 
\eqn\pois{\{\Psi^*(Y), \Psi(Y')\}_P = \d(Y-Y'), \quad 
\{\Psi(Y), \Psi^*(Y')\}_P = \d(Y-Y'), }
$$\{\Psi^*(Y), \Psi^*(Y')\}_P = 0, \quad 
\{\Psi(Y), \Psi(Y')\}_P = 0, $$
where $\{~,~\}_P$ denotes Poisson antibracket. However, because the constraint $\Psi^*-\Psi=0$ of \ccons\
is a second-class constraint, the Poisson antibracket needs to 
be modified to a Dirac antibracket in order that the antibracket
of $\Psi^*-\Psi$ with any other operator vanishes. As explained in \beringd, this modification follows the standard
Dirac procedure where the Dirac antibracket of operators $A$ and $B$ is defined by
\eqn\diracb{ \{A, B\} = \{A,B\}_P - \{A, C_I\}_P M^{IJ} \{C_J, B\}_P}
where $C_I$ are the second-class constraints and $M^{IJ}$ is the inverse of the matrix $\{C_I,C_J\}_P$.

For the constraints $C(Y)= \Psi^*(Y) - \Psi(Y)$, one finds that $\{C(Y),C(Y')\} = 
-2 \d(Y-Y')$ so that $M(Y,Y') = -{\half} \d(Y-Y')$. 
So the Dirac antibracket is given by
\eqn\diracd{\{\Psi^*(Y), \Psi(Y')\} = \half \d(Y-Y'), \quad
\{\Psi(Y), \Psi^*(Y')\} = \half \d(Y-Y'), }
$$\{\Psi(Y), \Psi(Y')\} = \half \d(Y-Y'), \quad
\{\Psi^*(Y), \Psi^*(Y')\} = \half \d(Y-Y').$$

As mentioned in the introduction, these Dirac antibrackets in string field theory
resemble the Dirac brackets of fermionic Weyl spinors in field theory. Just as
the $\langle \Phi Q\Phi \rangle$ kinetic term of string field theory is linear in 
$Q$, the $\int d^D x ~\psi^\a \g^m_{\a\b} \p_m \psi^\b$ kinetic term for a Weyl spinor
$\psi^\a$ is linear in ${\p\over{\p x^m}}$.
So canonical quantization implies that 
the canonical momenta $\psi^*_\a$ for
$\psi^\alpha$ satisfies the second-class
constraint $\psi^*_\alpha = \p L/\p(\p_0\psi^\a) = (\g^0 \psi)_\alpha$. 
The resulting Dirac brackets are 
\eqn\diracc{\{\psi^*_\a (x), \psi^\b (x')\} = \half \d_\a^\b \d^D(x-x'), \quad
\{\psi^\a (x), \psi^*_\b (x')\} = \half \d^\a_\b\d^D(x-x'), }
$$\{\psi^\a (x), \psi^\b (x')\} = \half (\g^0)^{\a\b}\d^D(x-x'), \quad
\{\psi^*_\a (x), \psi^*_\b (x') \} = \half \g^0_{\a\b}\d^D(x-x').$$

Comparing \diracd\ with \bPhione, one learns that the antibrackets of the constrained
BV formalism agree with the antibrackets of the conventional BV formalism where
$\Phi = {1\over{\sqrt {2}}}(\Psi + \Psi^*)$. So the nilpotent BV
transformation $\d_{BV}$ of \sbracket\ is unchanged in the constrained formalism. Using the
interpretation of $Q$ as a fermionic time derivative, it is a useful exercise
to check that the action $S$ is proportional to the Noether charge for this BV transformation.
As usual, the Noether charge can be constructed by computing the change in the action
when the constant parameter $\e$
of a global symmetry transformation is treated as a local parameter.
In classical mechanics, the change in the action 
is $\d S = \int dt  ({\p\over{\p t}}\e) f (q,{\p\over{\p t}} q)$ and
the Noether charge is defined by $f(q,{\p\over{\p t}}q)$. In the constrained BV description, the change in the action is
\eqn\change{\d S = \langle (Q\e) f(\Psi, Q\Psi)\rangle}
and the Noether charge is defined by $\langle f(\Psi, Q\Psi)\rangle$. 

In this case, the transformation is 
\eqn\noether{\d \Psi = \e \d_{BV}\Psi = 2\e (Q\Psi + \Psi\Psi)}
where $\e$ is treated as a local fermionic parameter.
Under this transformation, one finds that
\eqn\charge{\d S = \langle (\d\Psi) Q \Psi + \Psi Q \d\Psi + 2(\d\Psi) \Psi^2 \rangle}
$$ = \langle 4 (Q\e) (\Psi Q \Psi + {2\over 3} \Psi^3) \rangle$$
where $\langle Q(~) \rangle =0$ and $\langle \Psi^4 \rangle=0$ have been used. 
So the Noether charge is $4 \langle \Psi Q \Psi + {2\over 3}\Psi^3\rangle$
which is proportional to $S$ as claimed.

\newsec{Closed Bosonic String Field Theory}

\subsec{Conventional BV description}

In the conventional BV description of
closed bosonic string field theory, the action is \zw
\eqn\pclosedact{
S= \langle \Phi Q (c-\bar c)_0 \Phi \rangle + ...}
where $\Phi$ is a bosonic string field of unrestricted ghost number 
satisfying the constraint $(b-\bar b)_0\Phi = (T-\bar T)_0 \Phi =0$, $Q$ is the
sum of the left and right-moving BRST operators, $(b_0 ,c_0)$ and $(\bar b_0,\bar c_0)$ are the zero modes
of the left and right-moving Virasoro ghosts, $T$ and $\bar T$ are the
left and right-moving Virasoro constraints, $\langle ~\rangle$ denotes the measure factor on a sphere
of $+6$ ghost number, and $...$ denote the interaction terms.

In this conventional BV description, states of ghost number $\leq 2$ in $\Phi$ are identified as spacetime fields and states of
ghost number $>2$ are identified as spacetime antifields. 
So $\Phi$ splits as 
\eqn\splitcc{\Phi = \Psi + \Psi^*}
where $\Psi$ only includes states with ghost number $\leq 2$ and describes spacetime fields, and
$\Psi^*$ only includes states with ghost number $> 2$ and describes spacetime antifields. 
The BV antibracket of \brackone\ then implies that 
\eqn\brackPhicl{\{\Phi(Y), \Phi(Y')\} = (b-\bar b)_0 \d (Y-Y')}
where the factor of $(b_0-\bar b_0)$ comes from the constraint on $\Phi$.
Note that $\d(Y-Y')$ carries ghost number $+6$ and is proportional to 
\eqn\propclosed{(c- c')_{-1} (c-c')_0 (c-c')_1
(\bar c- \bar c')_{-1} (\bar c-\bar c')_0 (\bar c-\bar c')_1}
where $(c_1, c_0, c_{-1})$ and
$(\bar c_1,\bar c_0,\bar c_{-1})$ 
are the zero modes on a sphere of the left and right-moving Virasoro ghosts.
So $(b-\bar b)_0 \d(Y-Y')$ is proportional to
\eqn\propclosedtwo{(c+\bar c - c' -\bar c')_0 (c- c')_{-1} (c-c')_1
(\bar c- \bar c')_{-1} (\bar c-\bar c')_1 .}
Since $(b-\bar b)_0\d(Y-Y') = - (b' -\bar b')_0 \d(Y'-Y)$, the definition of \brackPhicl\ is
consistent with the fact that 
$\{\Phi(Y),\Phi(Y')\} = - \{\Phi(Y'), \Phi(Y)\}.$

Using the action $S$ of \pclosedact, one finds that
$S$ satisfies the antibracket
\eqn\scbracket{\{S,\Phi\} = Q\Phi + (b-\bar b)_0 (\Phi\Phi) + ...} 
which is interpreted as the BV transformation $\d_{BV}\Phi$. 
After including the appropriate interaction terms $...$, one finds that
\eqn\nilpotent{\d_{BV}\d_{BV}\Phi = \{S, \{S,\Phi\}\} =0. }
So the BV transformation is nilpotent as desired.

\subsec{Constrained BV description}

In the new constrained description, $\Phi$ of \pclosedact\ is  
replaced by $\Phi = (b-\bar b)_0 \Psi$ where $\Psi$ is a fermionic string field of unrestricted ghost number which
describes spacetime fields. In terms of $\Psi$, the action of \closedact\ is 
\eqn\closedtwoact{
S= \langle \Psi Q (b-\bar b)_0 \Psi \rangle + ...}
where the only constraint on $\Psi$ is the level-matching condition $(T_0-\bar T_0)\Psi=0$.
Note that the interaction term $...$ does not involve the $Q$ operator and all $\Psi$'s in the interaction term appear
in the combination $(b-\bar b)_0 \Psi$. So one can trivially show that \closedtwoact\ is invariant under the
gauge transformation
\eqn\closedg{\d\Psi = (b-\bar b)_0 \Lambda}
for arbitrary $\Lambda$ satisfying the level-matching condition.

Using the definition $\Psi^* = \p L/\p(Q\Psi)$ for the spacetime antifields, one finds that in addition to
the level-matching condition $(T_0-\bar T_0)\Psi^*=0$, $\Psi^*$ must satisfy the constraint
\eqn\closedanti{\Psi^* = \p L/\p(Q\Psi) = (b-\bar b)_0 \Psi.}
This constraint implies $(b-\bar b)_0\Psi^*=0$ which is a first-class constraint and generates
the gauge transformation of \closedg. 
So the constraint $\Psi^* - (b-\bar b)_0\Psi=0$ has a first-class and second class
piece and one has two options for modifying the Poisson antibracket
\eqn\closedP{\{\Psi^*(Y), \Psi(Y')\}_P = \d(Y-Y')}
into a Dirac antibracket.

The first option is to gauge-fix the invariance of \closedg\ which generates a new constraint. For example,
a convenient gauge choice is 
\eqn\convc{(c - \bar c)_0 \Psi=0.} In the presence of this new constraint, the constraint of \closedanti\
becomes completely second-class and one can follow the same Dirac procedure as in \diracb.

The second option is to not gauge-fix \closedg, in which case
the Dirac 
antibracket is only well-defined for operators which commute with the first-class constraints.
In other words, the Dirac bracket can only be defined for operators which are gauge-invariant with respect to \closedg.
In this second option, the matrix $M^{IJ}$ in \diracb\ is defined to be the inverse of the matrix $\{C_I,C_J\}$ where
$I$ and $J$ range only over the second-class constraints. The choice of how to split off these second-class constraints
from the first-class constraints does not lead to ambiguities since the operators $A$ and $B$ in \diracb\ are required to
have vanishing Poisson bracket with the first-class constraints.

Although both options are completely straightforward, 
only the second option will be discussed here. 
Gauge-invariant operators with respect to \gaugeclosed\ include the operators $(b-\bar b)_0 \Psi$ and $\Psi^*$, as well as the
action of \closedtwoact. Following the Dirac prescription of \diracb, one finds that
the Dirac antibracket is given by
\eqn\dirace{\{\Psi^*(Y), (b'-\bar b')_0 \Psi(Y')\} = 
\half (b-\bar b)_0 \d(Y-Y'), }
$$
\{(b-\bar b)_0 \Psi(Y), \Psi^*(Y')\} = 
\half (b-\bar b)_0 \d(Y-Y'), $$
$$\{(b-\bar b)_0 \Psi(Y), (b'-\bar b')_0 \Psi(Y')\} =
\half (b-\bar b)_0 \d(Y-Y'), $$
$$\{\Psi^*(Y), \Psi^*(Y')\} = \half (b-\bar b)_0\d(Y-Y').$$

If one defines the operator $\Phi$ in the conventional description by
$\Phi = {1\over{\sqrt 2}}[ (b-\bar b)_0 \Psi + \Psi^*]$, 
the antibrackets of \dirace\ reproduce the antibracket of \brackPhicl\
in the conventional description.
And since the action of \closedtwoact\  is equivalent to the action of \pclosedact , 
the BV transformations in the constrained description coming from $\d_{BV}\Phi = \{S, {1\over{\sqrt 2}}[(b-\bar b)_0 \Psi +\Psi^*]\}$
are the same as \scbracket\ in the conventional
description.

\newsec{Open Superstring Field Theory}

\subsec{ Classical action}

In thw WZW-like version of open superstring field theory, 
the classical action for the Neveu-Schwarz sector is \nb
\eqn\nsactiont{S = \langle (e^{-\Phi} Q e^\Phi)(e^{-\Phi}\eta_0 e^\Phi) +\int_0^1 dt
(e^{-t\Phi} \p_t e^{t\Phi})\{(e^{-t\Phi} Q e^{t\Phi}),(e^{-t\Phi}\eta_0 e^{t\Phi})\} \rangle}
where $\Phi$ is a string field in the ``large'' Hilbert space with zero picture and zero
ghost number, $Q$ is the worldsheet BRST operator, $\eta_0$ carries $-1$ picture and $+1$ ghost
number and is the zero mode coming from Friedan-Martinec-Shenker
fermionization of the $(\beta,\gamma)$ ghosts as $(\b=\p\xi e^{-\phi}, \g=\eta e^\phi)$,
and $\langle~\rangle$
is the standard measure factor in the ``large'' Hilbert space which carries $-1$ picture and $+2$ ghost number.
Note that $\langle \xi_0  c_{-1} c_0 c_1 \d(\g_{\half}) \d(\g_{-\half})\rangle$
is nonvanishing where $\d(\g_\half)\d(\g_{-\half}) = e^{-2\phi}$, 
$e^{n\phi}$ is defined to carry zero ghost number and $n$ picture,
and $(\eta,\xi)$ is defined to carry ghost number $(+1,-1)$ and picture $(-1,+1)$.
This classical action has the WZW-like gauge invariances 
\eqn\wzwgauge{\d (e^\Phi) = (Q \Lambda_{-1,0}) e^\Phi + e^\Phi (\eta_0 \Lambda_{-1,1})}
where $\Lambda_{g,p}$ are gauge parameters with ghost number $g$ and picture $p$.

It is interesting to point out that the same classical action of \nsactiont\ can be used to describe any critical $N=2$ open
string field theory by replacing $Q$ and $\eta_0$ with the corresponding operators $\int G^+$ and $\int\tilde G^+$ 
in the critical $N=2$ string. 
As explained in \nb\ref\review{N. Berkovits,
``Review of open superstring field theory'',
hep-th/0105230.}, 
this includes the $N=2$ string of \ref\vafa{H. Ooguri and C. Vafa, ``N=2 heterotic strings'', Nucl. Phys. B367 (1991) 83. } 
which describes $D=4$ self-dual Yang-Mills
theory, as well as the $N=2$ hybrid formalism of \ref\newd{N. Berkovits, ``New description of the superstring'',
hep-th/9604123.}
which describes the open superstring compactified on a Calabi-Yau three-fold.
So constructing the BV version of the open Neveu-Schwarz action of
\nsactiont\ automatically provides the BV action for these other theories. 

\subsec{ Conventional BV description}

As discussed in \free\us, the standard procedure for constructing the BV action is to introduce ghosts
for the gauge invariances of \wzwgauge\ and follow the Faddeev-Popov procedure. Because of the gauge-for-gauge invariances,
this gives rise to an infinite pyramid of bosonic ghosts-for-ghosts 
\eqn\gfg{\Phi}
$$\Phi_{-1,0} \quad\Phi_{-1,1}$$
$$\Phi_{-2,0} \quad\Phi_{-2,1} \quad \Phi_{-2,2}$$
$$...$$
where $\Phi_{g,p}$ carries ghost number $g$ and picture $p$ and comes from the gauge parameters of
\lingauge\ and \lingaugetwo. Furthermore, each of the bosonic fields $\Phi_{g,p}$
comes with a fermionic antifield $\Phi^*_{2-g,-p-1}$ with ghost number $2-g$ and picture $-p-1$. 

To quadratic order, the BV action can be easily determined from the structure of the linearized gauge invariances of
\lingauge\ and \lingaugetwo\ and one finds
\eqn\quadbv{S = \langle (Q\Phi)(\eta_0\Phi) + \Phi^*_{2,-1} (Q\Phi_{-1,0}+\eta_0\Phi_{-1,1})  }
$$+\Phi^*_{3,-1} (Q\Phi_{-2,0} + \eta_0\Phi_{-2,1}) + \Phi^*_{3,-2}(Q\Phi_{-2,1} + \eta_0\Phi_{-2,2}) + ... \rangle $$
Using the antibrackets of $\Phi^*_{2-g,-p-1}$ and $\Phi_{g,p}$, one 
can easily verify to
quadratic order that $\{S,S\}=0$.

The next step is to construct a nonlinear generalization of \quadbv\ where the term $\langle (Q\Phi)(\eta_0\Phi)\rangle$
is replaced by the WZW-like action of \nsactiont. Since the ghost fields $\Phi_{g,p}$ can have different picture from the classical
field $\Phi$, it is unclear if one should combine them into a single string field as was done in open and closed
bosonic string field theory. As will be described in \us, there are several approaches to constructing
this conventional BV action for open superstring field theory. However, a closed-form expression for the complete nonlinear BV action
has not yet been found using this conventional approach.

\subsec{Constrained BV description}

In the constrained BV description, the classical string field $\Phi$ in \nsactiont\ will be replaced
by a bosonic string field $\Psi$ of zero picture and unrestricted ghost number which will describe the spacetime fields.
So the BV action is 
\eqn\nsactionu{S = \langle (e^{-\Psi} Q e^\Psi)(e^{-\Psi}\eta_0 e^\Psi) +\int_0^1 dt
(e^{-t\Psi} \p_t e^{t\Psi})\{(e^{-t\Psi} Q e^{t\Psi}),(e^{-t\Psi}\eta_0 e^{t\Psi})\} \rangle.}
One then introduces a fermionic string field $\Psi^*$ of $-1$ picture and unrestricted ghost number to describe the spacetime
antifields and imposes the constraint that 
\eqn\supercons{\Psi^* = \p L/\p(Q\Psi)}
where $L$ is the WZW-like Lagrangian of \nsactionu. To be more explicit, the exponentials in the WZW-like action can
be expanded in a power series to give
\eqn\power{S = \sum_{M,N=0}^{\infty} {2\over{M! N! (M+N+1)(M+N+2)}} (-1)^N \langle (Q\Psi) \Psi^M (\eta_0\Psi) \Psi^N \rangle.}
So the constraint of \supercons\ is
\eqn\sucons{\Psi^* = \sum_{M,N=0}^{\infty} {2\over{M! N! (M+N+1)(M+N+2)}} (-1)^N  \Psi^M (\eta_0\Psi) \Psi^N .}

To quadratic order, the BV action is $S=\langle (Q\Psi)(\eta_0\Psi)\rangle$ and the BV constraint is 
$\Psi^* = \eta_0\Psi$. This closely resembles the quadratic action \closedtwoact\
and constraint \closedanti\ of closed bosonic string field theory
where $\eta_0$ is replaced by $(b-\bar b)_0$. However, unlike in closed bosonic string field theory
where the linearized gauge invariance $\d\Psi = (b-\bar b)_0\Lambda$ is unaffected by interactions,
the linearized gauge invariance $\d\Psi= \eta_0\Lambda$ of open superstring field theory generalizes
to the nonlinear gauge invariance 
\eqn\nonling{\d(e^\Psi) = e^\Psi (\eta_0 \Lambda).}
So the linearized constraint
$\Psi^*= \eta_0\Psi$ is also generalized to the nonlinear constraint of \sucons.

The constraint of \sucons\ contains both a first-class piece and a second-class piece
where the first-class piece generates the gauge invariance of \nonling. So as discussed in the 
previous section, one option for defining Dirac brackets is to gauge-fix \nonling\ and turn the first-class constraints into 
second-class constraints. Alternatively, one can define Dirac brackets of operators which are gauge-invariant with respect to \nonling.
We will follow the second option here, but will later discuss the first option when we compare the constrained BV approach with 
Witten's cubic version of open superstring field theory.

Gauge-invariant operators with respect to \nonling\ include the operator $J= (\eta_0 e^\Psi) e^{-\Psi}$ as well as the action
$S$ of \power. The Poisson antibrackets of $\Psi$ and $\Psi^*$ are defined as usual by
\eqn\poist{\{\Psi^*(Y), \Psi(Y')\}_P = \d(Y-Y'), \quad 
\{\Psi(Y), \Psi^*(Y')\}_P = \d(Y-Y'), }
$$\{\Psi^*(Y), \Psi^*(Y')\}_P = 0, \quad 
\{\Psi(Y), \Psi(Y')\}_P = 0. $$
To compute the Dirac antibrackets of $J$ and $S$, it is convenient to first compute the Dirac antibrackets at the linearized level
and then use the nonlinear gauge invariance to deduce the nonlinear antibrackets.

At the linearized level, the constraint is $\Psi^* - \eta_0\Psi=0$ and the linearized Dirac antibrackets are 
\eqn\diract{\{\Psi^*(Y), \eta_0\Psi(Y')\} = 
\half\eta_0\d(Y-Y'), \quad
\{\eta_0\Psi(Y), \Psi^*(Y')\} = 
\half\eta_0\d(Y-Y'), }
$$\{\Psi^*(Y), \Psi^*(Y')\} 
=\half\eta_0\d(Y-Y'), \quad
\{\eta_0\Psi(Y), \eta_0\Psi(Y')\}_P = 
\half\eta_0\d(Y-Y'). $$
Note that $\d(Y-Y')$ is defined in the large Hilbert space and 
is proportional to 
\eqn\ysuper{(c-c')_1 (c-c')_0 (c-c')_{-1} (\xi-\xi')_0 \d(\g_{\half}-\g'_{\half}) \d (\g_{-\half}-\g'_{-\half})}
where $(\g_\half, \g_{-\half})$ are the zero modes of the bosonic $\g$ ghost on a disk.
So $\eta_0\d(Y-Y')$ is proportional to
\eqn\zsuper{(c-c')_1 (c-c')_0 (c-c')_{-1} \d(\g_{\half}-\g'_{\half}) \d (\g_{-\half}-\g'_{-\half})}
and satisfies $\eta_0\d(Y-Y') = - \eta'_0\d(Y-Y')$.
Using these linearized antibrackets and the quadratic action $S=\langle (Q\Psi)(\eta_0\Psi)\rangle$, 
the antibracket of $S$ with $\eta_0\Psi$ is easily computed to be 
\eqn\linab{\{S,\eta_0\Psi\} = Q(\eta_0\Psi).}

At the linearized level, $J=(\eta_0 e^\Psi) e^{-\Psi} $ reduces to $J=\eta_0\Psi$ and satisfies the antibracket
$\{J(Y),J(Y')\} = \half \eta_0 \d(Y-Y')$, i.e.
\eqn\linanti{\{J^K(Y), J^L(Y')\} = \half h^{KL} \eta_0\d(Y-Y')}
where $J = J^K T_K$, $T_K$ are Lie algebra generators
coming from the Chan-Paton factors, and $h^{KL}$ is the Lie algebra metric.
But $\eta_0 J = (\eta_0 e^\Psi) e^{-\Psi} (\eta_0 e^\Psi) e^{-\Psi} $
implies that 
\eqn\etaJ{\eta_0 J^K = f^K_{LM} J^L J^M}
where $f^K_{LM}$ are the Lie algebra structure constants.
This is inconsistent with the linearized antibracket of \linanti\ since the right-hand side of \linanti\ is annihilated by $\eta_0$.
Fortunately, one can modify \linanti\ to the nonlinear antibracket 
\eqn\nonlinanti{\{J^K(Y), J^L(Y')\} = \half h^{KL} \eta_0\d(Y-Y') + f^{KL}_M J^M \d(Y-Y')}
which is consistent with \etaJ\ after using the Jacobi identity $f^K_{L[M} f^L_{NP]}=0$ for the structure constants.
Furthermore, \nonlinanti\ is the unique gauge-invariant modification of \linanti\ which is
consistent with \etaJ. As pointed out in the introduction, the structure of \nonlinanti\ closely resembles the OPE
of holomorphic Kac-Moody currents $J = (\p g) g^{-1}$ in a two-dimensional
WZW model which is
\eqn\OPE{J^K(z) J^L(z') \to (z-z')^{-2} h^{KL} + (z-z')^{-1} f^{KL}_M J^M(z).}

Starting with the linearized antibracket of $S$ with $J$ in \linab, one can also use nonlinear gauge invariance to deduce
that the unique gauge-invariant option for the nonlinear antibracket is 
\eqn\nonlinS{\d_{BV} J^K = \{S, J^K\} = Q J^K.} 
Since $Q^2=0$, this immediately implies that $\d_{BV}$ is nilpotent when acting on $J^K$. 
Note that
\nonlinS\ implies that
\eqn\nongs{\d_{BV} e^\Psi  = Q (e^\Psi) + e^\Psi (\eta_0\Lambda)}
for some $\Lambda$. Since gauge-invariant operators do not 
transform under $\d e^\Psi = e^\Psi (\eta_0\Lambda)$, one
learns from \nongs\ that $\d_{BV} ({\cal O}) = Q{\cal O}$
for any gauge-invariant operator ${\cal O}$ constructed from $\Psi$.

Since $\d e^\Psi = \e \d_{BV} e^\Psi =  \e
(Q e^\Psi + e^\Psi (\eta_0\Lambda))$
is a global symmetry of the action, one can compute the Noether charge for this symmetry
using the same method as in \change. Using the fact that all terms in the Lagrangian
$L$ of \nsactionu\  have one $Q$ operator,
one finds that 
\eqn\nosup{\d S = \langle \e Q L + (Q\e) L \rangle = 2 \langle (Q\e) L\rangle}
where terms involving $(\eta_0 \e)$ have been dropped. So using the Noether method described in
\change, the Noether charge for the BV transformation is proportional to  $S=\langle L\rangle$ as expected.

Because the nonlinear Dirac brackets of \nonlinanti\ and \nonlinS\ were deduced from nonlinear gauge
invariance and were not explicitly computed, it is useful
to collect additional evidence that the constrained BV description for open superstring field theory is consistent.
As will now be discussed, two additional pieces
of evidence come from comparison of the constrained BV description
with the conventional BV descriptions of WZW-like and cubic open superstring field theory. 

\subsec{Comparison with conventional BV descriptions}

As discussed in \free\us, the quadratic terms in the conventional BV action for
WZW-like open superstring field theory are given by 
\quadbv\ where $\Phi_{g,p}$ is a bosonic string field with ghost-number $g$ and picture $p$,
and $\Phi^*_{g,p}$ is a fermionic string antifield with ghost-number $g$ and picture $p$.
This action can be conveniently expressed as 
\eqn\conven{S = \langle \Phi_0 Q\eta\Phi_0 + \sum_{p\geq 0} \Phi_{-p-1}^* (Q\Phi_p + \eta\Phi_{p+1})\rangle }
where for $p\geq 0$,
\eqn\Phidef{\Phi_p \equiv \sum_{g\leq -p} \Phi_{g,p}, \quad \Phi^*_{-p-1} = \sum_{g\geq 2+p} \Phi^*_{g,-p-1} ,}
and only terms of total ghost number $+2$ contribute to $S$.

It will now be shown that after partially gauge-fixing and solving for auxiliary fields, the conventional
BV action of \conven\
reduces to the quadratic term in the constrained BV action of \nsactionu. If the complete nonlinear version
of the conventional
BV action could be constructed, it seems reasonable to conjecture that a similar gauge-fixing procedure would
reduce this action to the complete nonlinear action of \nsactionu.

The first step is to note that the conventional BV action of \conven\ 
contains the linearized gauge invariances
\eqn\lineari{
\d\Phi_{p-1} = \eta_0\L_p, \quad \d\Phi_p = Q\L_p}
for $p\geq 1$ where $\L_p$ includes states of ghost number $\leq -p$, and also contains
the linearized gauge invariances
\eqn\linearit{\d\Phi^*_{-p-1} = \eta_0\L_{-p}, \quad \d\Phi^*_{-p} = Q\L_{-p}}
for $p\geq 1$ where $\L_{-p}$ includes states of ghost number $\geq 1+p$.
Using these gauge invariances, one can gauge
$\xi_0 \Phi_p =0$ for $p\geq 0$ and
$\xi_0 \Phi^*_p =0$ for $p\leq -2$. So only $\Phi^*_{-1}$ cannot be
gauged to satisfy $\xi_0 \Phi^*_{-1}=0$.

In this gauge, one can easily verify that the equations of motion
$Q\Phi_p + \eta_0 \Phi_{p+1} =0$ imply that $\Phi_p$ are auxiliary
fields for all $p>0$. In other words, up to gauge transformations,
all $\Phi_p$'s can be solved onshell in terms of $\Phi_0$. Furthermore,
the equations of motion $Q\Phi^*_{-p-2} + \eta_0\Phi^*_{-p-1}=0$ imply
that $\Phi^*_{-p-1}$ are auxiliary fields for all $p>0$. These equations
imply that all $\Phi^*_p$'s for $p< -1$ can be gauged to zero, and
that $\eta_0 \Phi^*_{-1}=0$.

So the only fields which are not auxiliary are $\Phi_0$ and $\Phi^*_{-1}$,
and after solving for the auxiliary fields, the action for these
remaining fields is
$S = \langle \Phi_0 Q\eta_0\Phi_0 + \Phi^*_{-1} Q \Phi_0 \rangle $
where $\eta_0\Phi^*_{-1}$ is constrained to vanish.
The constraint $\eta_0\Phi^*_{-1}=0$ can be solved as $\Phi^*_{-1} = \eta_0\Sigma_0$
for some $\Sigma_0$ at picture zero which includes ghost numbers $\geq 1$.
If one now defines
\eqn\defnewf{\Psi = \Phi_0 +{1\over 2}
 \Sigma_0 , \quad \Psi^* = \eta_0\Phi_0 + {1\over 2} \Phi^*_{-1},}
$\Psi$ and $\Psi^*$ have unrestricted ghost number and satisfy the constraint
$\Psi^*=\eta_0\Psi$. Furthermore, the action of \conven\
reduces to $S= \langle \Psi Q \eta_0 \Psi\rangle$ which is
the quadratic term in \nsactionu. So it has been shown
at the quadratic level
that the constrained BV action reproduces a partially gauge-fixed
version of the conventional BV action.

It will now be shown that it is also possible to relate the constrained BV
action of \nsactionu\ with the conventional BV action of cubic open superstring field theory \ref\witcub{E. Witten,
``Interacting Field Theory of Open Superstrings'',
Nucl.Phys. B276 (1986) 291.}\ref\preit{C.R. Preitschopf, C.B. Thorn and S. Yost, ``Superstring
field theory'', Nucl. Phys. B337 (1990) 363. }\ref\aref{
 I. Ya. Aref’eva, P.B. Medvedev and A.P. Zubarev, 
`'New Representation For String Field Solves The Consistency Problem For Open Superstring Field Theory'', Nucl. Phys. B341 
(1990) 464.}. 
Although the cubic version of open
superstring field theory is singular
because of midpoint insertions of picture-changing operators \ref\wendt{C. Wendt,
``Scattering Amplitudes And Contact Interactions In Witten's Superstring Field Theory'',
Nucl.Phys. B314 (1989) 209.}, its conventional BV description is similar to that
of open bosonic string field theory. We will only explicitly compare with Witten's cubic version of open
superstring field theory \witcub, but it should be possible to also compare with other versions of cubic open
superstring field theory \preit\aref\ using the methods discussed in \ref\kroyter{M. Kroyter,
``Democratic Superstring Field Theory: Gauge Fixing'',
JHEP 1103 (2011) 081,
arXiv:1010.1662 [hep-th].}.

Witten's cubic action for Neveu-Schwarz superstring field theory is \witcub 
\eqn\wit{S = \langle V Q V + {2\over 3}\{Q,\xi({\pi\over 2})\} VVV \rangle_S}
where the classical string field $V$ is defined in the small
Hilbert space at $-1$ picture and $+1$ ghost number, 
$\{Q,\xi({\pi\over 2})\}$ is the picture-raising operator inserted at the string midpoint,
and $\langle ~\rangle_S$ is defined in the small Hilbert space without the $\xi$ zero mode.
Although this cubic action has contact-term problems because of colliding picture-changing
operators \wendt, one can easily define a BV version of the action by allowing $V$ to have unrestricted
ghost number. The BV antibracket is given by
\eqn\antibv{\{V(Y), V(Y')\} = \eta_0 \d(Y-Y') }
where the factor of $\eta_0$ comes from the measure factor $\langle ~\rangle_S$ being defined
in the small Hilbert space. And the BV transformation of $V$ is given by
\eqn\bvV{\d_{BV} V =\{S, V\} = QV + \{Q,\xi({\pi\over 2})\} V V,}
which is nilpotent up to contact-term problems.

To relate this cubic action with the constrained BV description of the WZW-like action,
use the nonlinear gauge invariance of \nonling\ to gauge-fix
\eqn\conssuper{\xi({\pi\over 2}) \Psi =0.} 
Although this gauge-fixing is singular since it involves insertions at the string midpoint,
it will allow a comparison of \nsactionu\ with Witten's cubic action.
After including the constraint of \conssuper, the constraint of \sucons\
becomes second-class and one can use the standard Dirac procedure of \diracb\ to define
the Dirac antibracket. Furthermore, the constraint of \conssuper\ implies that
\eqn\xiv{\Psi = \xi({\pi\over 2}) V}
for some $V$ in the small Hilbert space.
Plugging \xiv\ into the action of \nsactionu\ and using that
$\xi({\pi\over 2})\xi({\pi\over 2}) =0$, one obtains
\eqn\wittwo{S = \langle \xi({\pi\over 2}) (V Q V + {2\over 3}
\{Q,\xi({\pi\over 2})\} V VV) \rangle.}
But this is equal to \wit\ after removing the midpoint insertion $\xi({\pi\over 2})$ which
converts the large Hilbert space measure factor $\langle~\rangle$ into the small Hilbert
space measure factor $\langle ~\rangle_S$.

Finally, one can use the procedure of \diracb\ to compute the Dirac antibracket and one finds
the same antibrackets as \antibv\ and \bvV. So in the singular gauge of \conssuper, the constrained BV
description of open superstring field theory coincides with the conventional BV version of Witten's cubic open
superstring field theory.

\vskip 15pt

{\bf Acknowledgements:}
I would 
like to thank Klaus Bering, Michael Kroyter, Andrei Mikhailov, Yuji Okawa, Martin Schnabl, Warren Siegel, Shingo Torii and
Barton Zwiebach for 
useful discussions, 
CNPq grant 300256/94-9
and FAPESP grants 09/50639-2 and 11/11973-4 for partial financial support, and 
MIT and the organizers of String Field Theory 2011 for their hospitality
where part of this work was done.

\listrefs

\end